\documentclass{PoS}

\usepackage{amsmath,amssymb}
\usepackage{amsfonts}
\usepackage[font=small]{caption}
\usepackage{url}
\usepackage{relsize}
\usepackage[utf8]{inputenc}
\DeclareMathAlphabet{\mathcal}{OMS}{cmsy}{m}{n}

\let\OLDthebibliography\thebibliography
\renewcommand\thebibliography[1]{
  \OLDthebibliography{#1}
  \setlength{\parskip}{0pt}
  \setlength{\itemsep}{0pt plus 0.3ex}
}

\newcommand{\tij}[1]{\hat{T}_{#1}}
\newcommand{\tnom}[1]{\hat{T}_{#1}^{0}}

\newcommand{\Fnom}[1]{\hat{R}_{#1}^{0}}
\newcommand{\norm}{\mathcal{K}}

\title{The Cherenkov transparency coefficient for the atmospheric monitoring and array calibration \\ at the Cherenkov Telescope Array South}

\ShortTitle{The Cherenkov transparency coefficient for the CTA South}

\author{\speaker{Stanislav Stefanik}, Dalibor Nosek for the CTA Consortium\thanks{https://www.cta-observatory.org. For collaboration list see PoS(ICRC2019)1177.}\\
        Charles University, Faculty of Mathematics and Physics\\
V Holesovickach 2, 180 00 Prague, Czech Republic\\
        E-mail: \email{stefanik@ipnp.troja.mff.cuni.cz}}

\abstract{
Reconstruction of energies of very-high-energy $\gamma$--rays observed by imaging atmospheric Cherenkov telescopes is affected by changes in the atmospheric conditions and the performance of telescope components. 
Reliable calibration schemes aimed at these effects are necessary for the forthcoming Cherenkov Telescope Array (CTA) to achieve its goals on the maximally allowed systematic uncertainty of the global energy scale. 
A possible means of estimating the atmospheric attenuation of Cherenkov light is the method of the Cherenkov transparency coefficient (CTC). 
The CTC is calculated using the telescope detection rates, dominated by the steady cosmic ray background, while properly correcting for the hardware and observational conditions. 
The coefficient can also be used to relatively calibrate the optical throughput of telescopes on the assumption of homogeneous atmospheric transparency above the array. 
Using Monte Carlo simulations, we investigate here the potential of the CTC method for the atmospheric monitoring and telescope cross-calibration at the CTA array in the southern hemisphere. 
We focus on the feasibility of the method for the array of telescopes of three sizes in different observation configurations and under various levels of atmospheric attenuation.
}

\FullConference{36th International Cosmic Ray Conference -ICRC2019-\\
		July 24th - August 1st, 2019\\
		Madison, WI, U.S.A.}

\begin{document}

% ******************************************************************************************
% INTRODUCTION
% ******************************************************************************************

\section{Introduction}
\label{Sec:Introduction}

The Cherenkov Telescope Array (CTA) is a future ground-based observatory of very-high-energetic (VHE) $\gamma$--rays of cosmic origin~\cite{Acharya:2013,Mazin:2019}. 
Two arrays of imaging atmospheric Cherenkov telescopes (IACTs) will detect Cherenkov radiation from air showers initiated by cosmic particles. 
Improved energy reconstruction compared to the current systems will require reliable calibration of the telescope arrays as well as the atmosphere as an inseparable part of the detector~\cite{Maccarone:2017, Ebr:2017}.

A possible calibration approach for the CTA is the method of the Cherenkov transparency coefficient (CTC) which monitors the transparency ($T$) of the atmosphere to Cherenkov light from air showers~\cite{Hahn:2014}. 
It uses telescope data and the assumption that the number of air showers recorded per unit time (trigger rate) is sensitive to the atmospheric transmission.
The method can also be used to assess the telescope optical throughput by utilizing the hardware dependence of the trigger~rate. 

The original method of the CTC cannot be easily applied without loss of precision to arrays of tens of telescopes with diverse designs. %~\cite{Stefanik:2017}.
An extension of the CTC has been recently proposed as a feasible means of the atmospheric and array calibration at the CTA North observatory~\cite{Stefanik:2019a}. 
Here, we continue with this work and present a preliminary feasibility study for the implementation of the extended CTC in the CTA array in the southern hemisphere~\cite{Maier:2019}. 
The viability of the CTC is examined in the context of different characteristics of the southern observatory compared to the northern one: the size of the local geomagnetic field and the number of distinct telescope designs.

The study uses Monte Carlo (MC) simulations of proton-initiated air showers observed by the large- (LST), medium- (MST) and small-sized (SST) telescopes to be deployed at the CTA South. 
In particular, the feasibility of the CTC for the SSTs is studied here for the first time. 
We assume the SST-1M design~\cite{Heller:2019}, one of the initial prototypes of the CTA small telescopes. 

The proceedings paper is structured as follows: an alternative approach to the original CTC calculation~\cite{Stefanik:2019a} is outlined in Section~\ref{Sec:CTC}. 
Section~\ref{Sec:Application} presents results from our MC study on the feasibility of the CTC for atmospheric monitoring (\ref{Sec:Monitoring}) and array calibration (\ref{Sec:Cross_calibration}) at the CTA South. 
Conclusions and further outlook are given in Section~\ref{Sec:Conclusions}.

% ******************************************************************************************
% CHERENKOV TRANSPARENCY COEFFICIENT
% ******************************************************************************************
\vspace{-0.1cm}
\section{Cherenkov transparency coefficient}
\label{Sec:CTC}

In a first order approximation, trigger rates of IACTs are given by the integral flux of charged cosmic rays weighted by the effective area of telescopes~\cite{Hahn:2014}. 
Since the flux of cosmic rays is assumed to be constant in time, variations of the trigger rates are indicative of changes in the atmospheric transparency, provided other phenomena affecting the rates are accounted for. 
Complex dependencies of trigger rates in large telescope arrays are included in the updated method of the CTC~\cite{Stefanik:2019a}. 
We differentiate two implementations of the method:
\begin{itemize}
\item {\em Atmospheric monitoring}.
An estimate of the atmospheric transparency to Cherenkov photons, given by the aerosol optical depth ($\tau$), is obtained as~\cite{Stefanik:2019a}:
\begin{equation}
    \label{Eq:CTC}
    \tij{} = \frac{1}{P} \mathlarger{\sum}_{ \substack{i = 1 \\ i < j} }^{N} \tij{ij} (\tau) = \frac{1}{P \cdot \norm}  \mathlarger{\sum}_{ \substack{i = 1 \\ i < j} }^{N} \left( \frac{R_{ij} (\tau, \mathcal{O}, \varepsilon_{i}, \varepsilon_{j})}{\varepsilon_{i} \cdot \varepsilon_{j} \cdot \Fnom{ij} (\mathcal{O})} \right)^{\frac{1}{\gamma}},
    \vspace{-0.2cm}
\end{equation}
\noindent
where $\hat{T}$ denotes the Cherenkov transparency coefficient, $\tij{ij}$ is the coefficient calculated for a pair of telescopes with identifiers $i,j$ and optical throughput efficiencies $\varepsilon_{i}, \varepsilon_{j}$. 
$R_{ij}$ is the rate of events triggering telescopes $i,j$ in coincidence, $\Fnom{ij}$ is the estimate of the pairwise trigger rate recovered from MC simulations for observational conditions $\mathcal{O}$ and $\varepsilon_{i} = \varepsilon_{j} = 1$. 
$N$ is the number of active telescopes during the observation, $P$ the number of telescope pairs and $\mathcal{K}$ is the normalization for the reference atmospheric profile. 
Index $\gamma$ derives from the dependence of the flux of registered cosmic particles on their energy. 
We assume $\gamma = \Gamma - 1$, where $\Gamma = 2.7$, as the trigger rates of IACTs are dominated by air showers induced by charged cosmic rays with differential flux $J(E) \propto E^{-\Gamma}$~\cite{Sanuki:2000}. 

\item {\em Array calibration}. 
Optical throughput efficiencies $\varepsilon_{i}$ in~Eq.\eqref{Eq:CTC} need to be provided by an independent calibration procedure. 
If the efficiencies are not known or have to be cross-checked, the relationship between the CTC and the telescope hardware can be exploited for relative calibration of telescope responses. 
Provided the atmospheric transparency in a given pointing direction of telescopes is stable across the array ($\tij{ij} = T, \forall i,j$), the inter-calibration of telescopes of the same type can be achieved by minimizing the objective function~\cite{Stefanik:2019a}
\vspace{-0.2cm}
\begin{equation}
    \label{Eq:chi2}
    F(\varepsilon, T) = \mathlarger{\sum}_{ \substack{i = 1 \\ i < j} }^{N}{ \frac{ \left( \tnom{ij} - \left( \varepsilon_{i} \cdot \varepsilon_{j} \right)^{\frac{1}{\gamma}} \cdot T \right)^{2} }{\sigma^{2}_{ij}} },
    \vspace{-0.4cm}
\end{equation}
\noindent
where $\tnom{ij} = \left( R_{ij} / \Fnom{ij} \right)^{1/\gamma}$, $\sigma_{ij}$ are the uncertainties of estimates $\tnom{ij}$ and $\varepsilon_{i}$, $T$ are free parameters. 
Assuming the value of $T$ is not known beforehand, the scale of optimized efficiencies $\hat{\varepsilon}$ is not fixed. 
The optical throughput of one telescope is arbitrarily chosen as a reference one ($\varepsilon_{\mathrm{R}}$) and is not varied. 
The efficiency estimates of other telescopes and the pairwise transparency estimates are expressed relatively as $\hat{\varepsilon}_{i} / \varepsilon_{\mathrm{R}}$ and $q_{ij} = \tij{ij} \cdot \varepsilon_{\mathrm{R}}^{-2 / \gamma}$, respectively. 
The cross-calibration of optical throughput between two sub-systems of different telescope types $A$ and $B$ is then performed by relating the reference efficiencies as $\varepsilon_{\mathrm{R}}^{A} = \left[ \langle q^{A} \rangle / \langle q^{B} \rangle \right]^{\gamma / 2} \cdot \varepsilon_{\mathrm{R}}^{B}$. 
\end{itemize}

\begin{figure}[!t]
	\begin{center}
		\includegraphics[width=0.44\columnwidth]{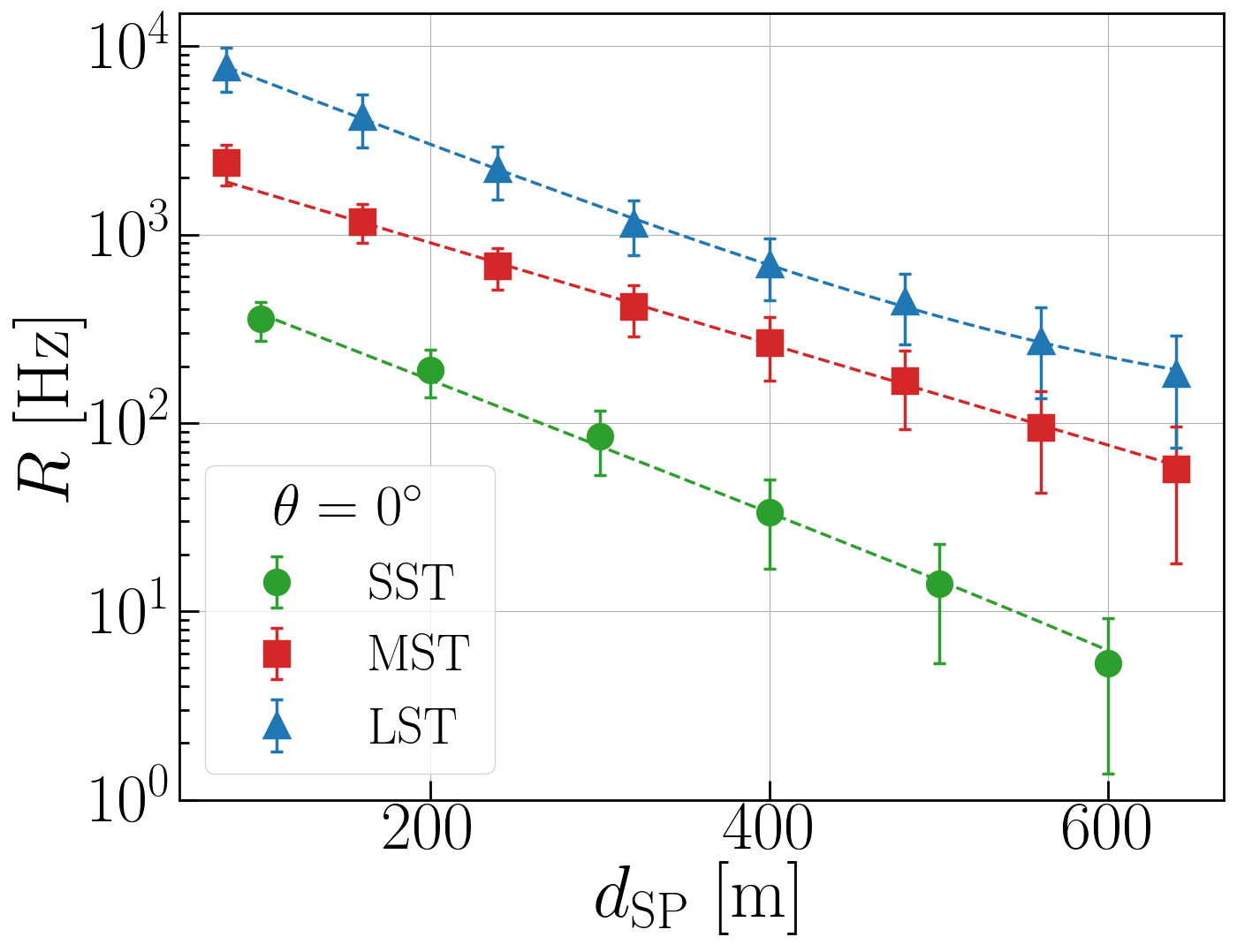}
		\hspace{0.05\columnwidth}
		\includegraphics[width=0.44\columnwidth]{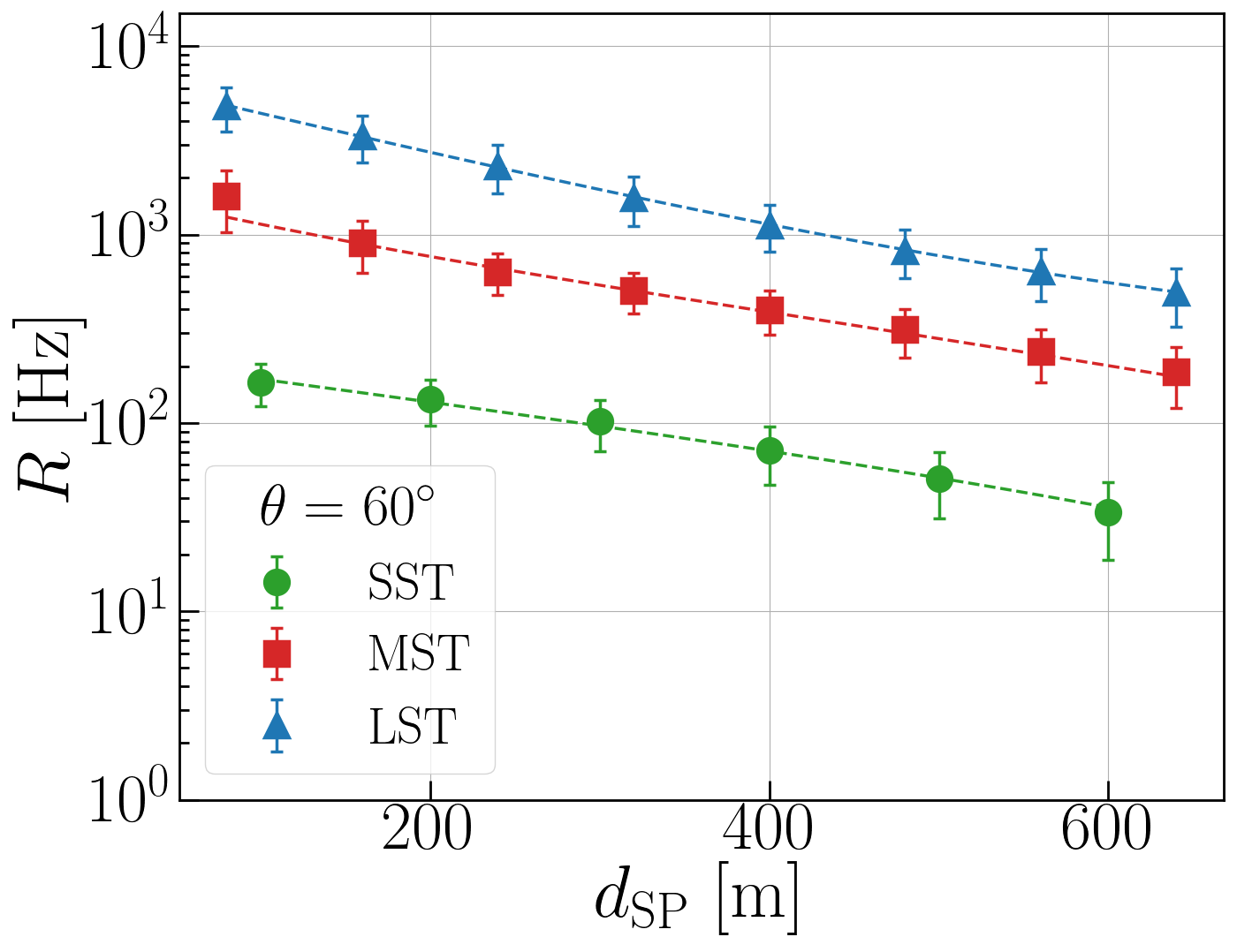}
		\caption{
			Simulated trigger rates are shown for pairs of telescopes of the same type as a function of the telescope separation in the shower plane~($d_{\mathrm{SP}}$, see top panel in Fig.~\protect\ref{Fig:Bfield}) for zenith angles $\theta = 0^{\circ}$ (left) and $60^{\circ}$ (right). 
			Colours denote telescope classes. 
			The magnetic field in the simulations was assumed to be $B \approx 0$.
		}
		\label{Fig:Geometry}
	\end{center}
	\vspace{-0.7cm}
\end{figure}

Trigger rates comprising events detected by two telescopes in coincidence are used in order to mitigate the contribution of accidental triggers. 
Pairwise transparency estimates in Eq.\eqref{Eq:CTC}, and thus the CTC, are made hardware independent by accounting for the optical throughput of both instruments ($\varepsilon_{i}, \varepsilon_{j}$) constituting a telescope pair ($i,j$). 
Telescope efficiencies are expressed relatively as a fraction of the nominal optical throughput, i.e.~$\varepsilon \in [0,1]$.

\begin{figure}[!t]
	\begin{center}
		\includegraphics[width=0.25\columnwidth]{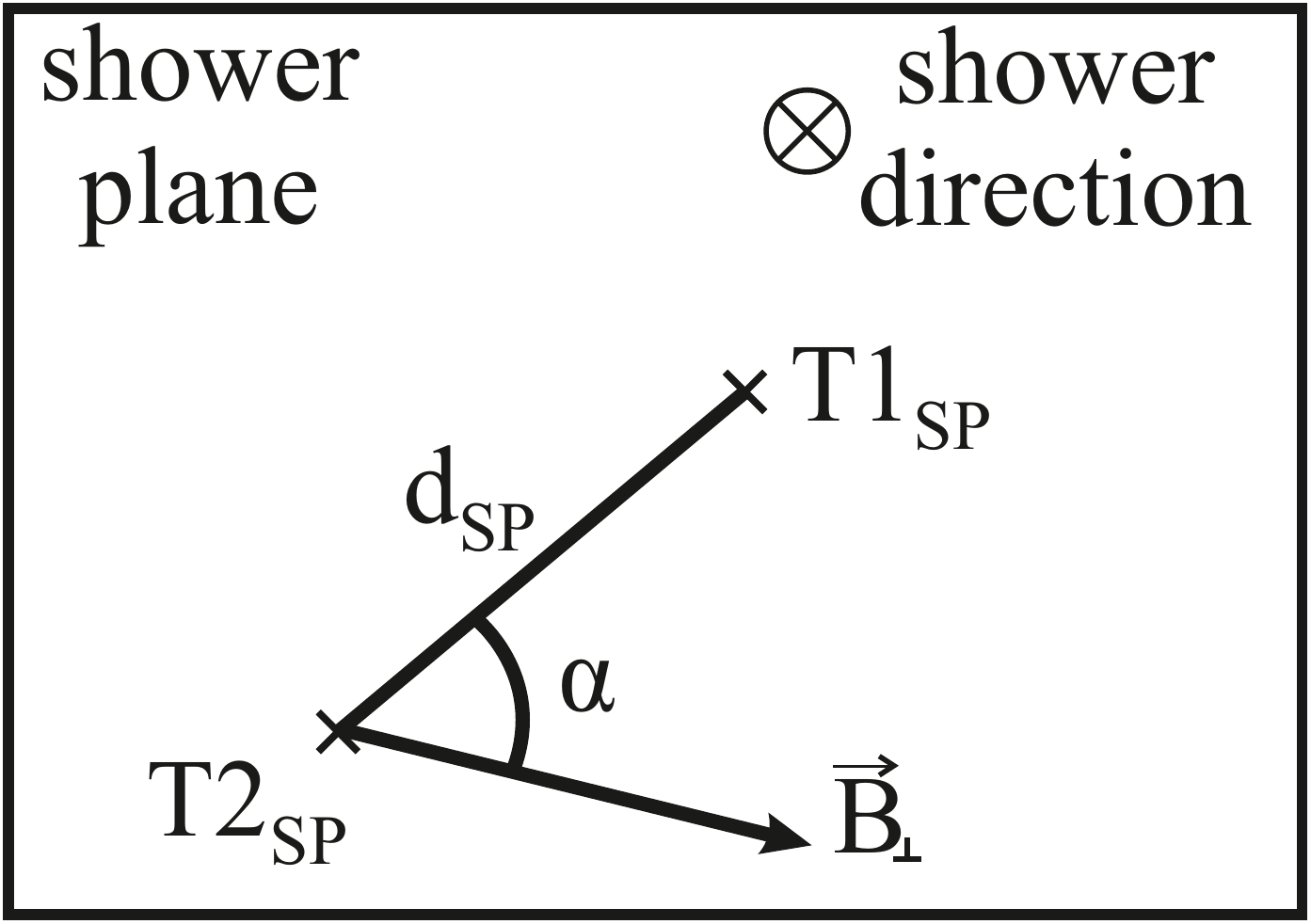} \\
		\vspace{0.2cm}
		\includegraphics[width=0.44\columnwidth]{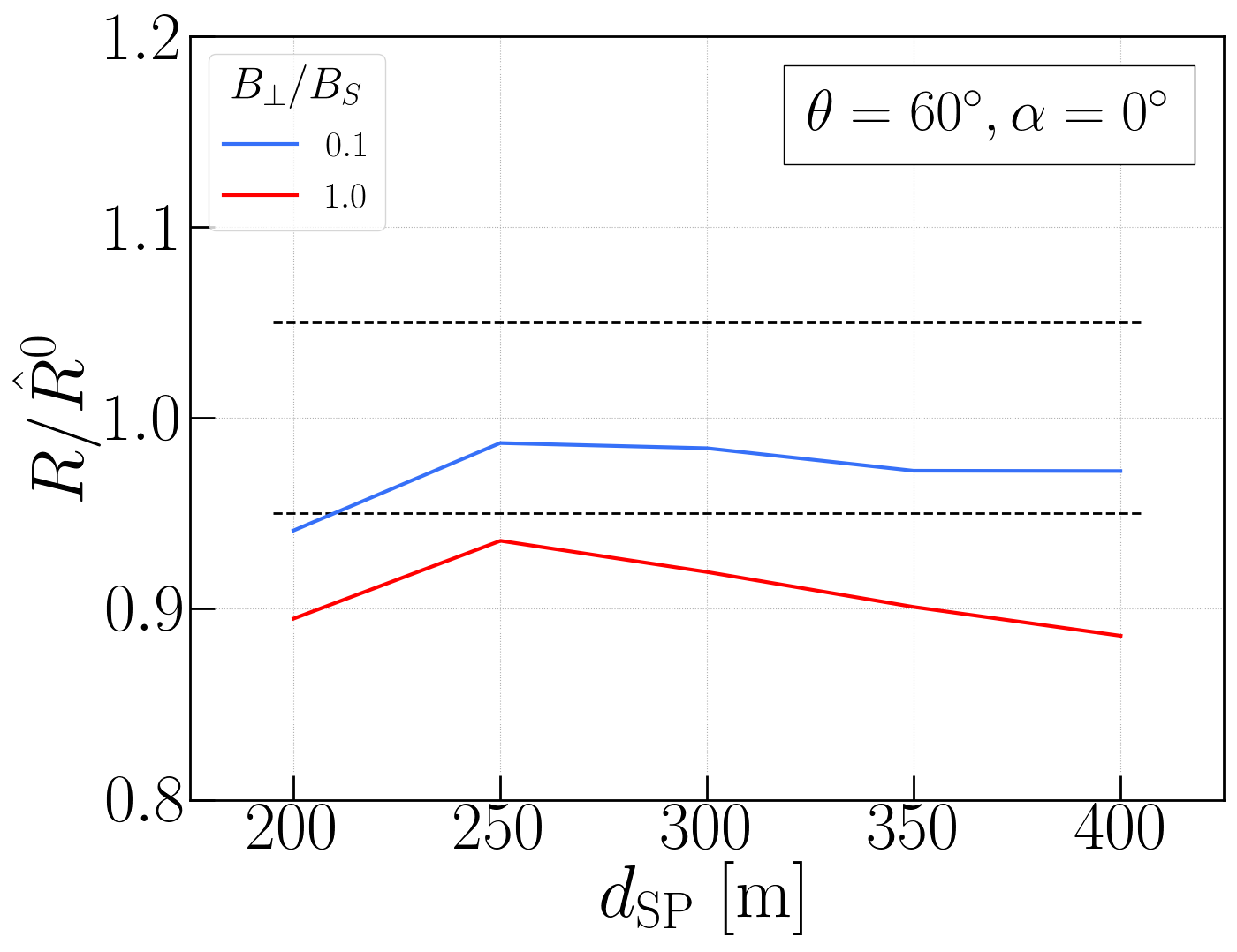}			
		\hspace{0.05\columnwidth}
		\includegraphics[width=0.44\columnwidth]{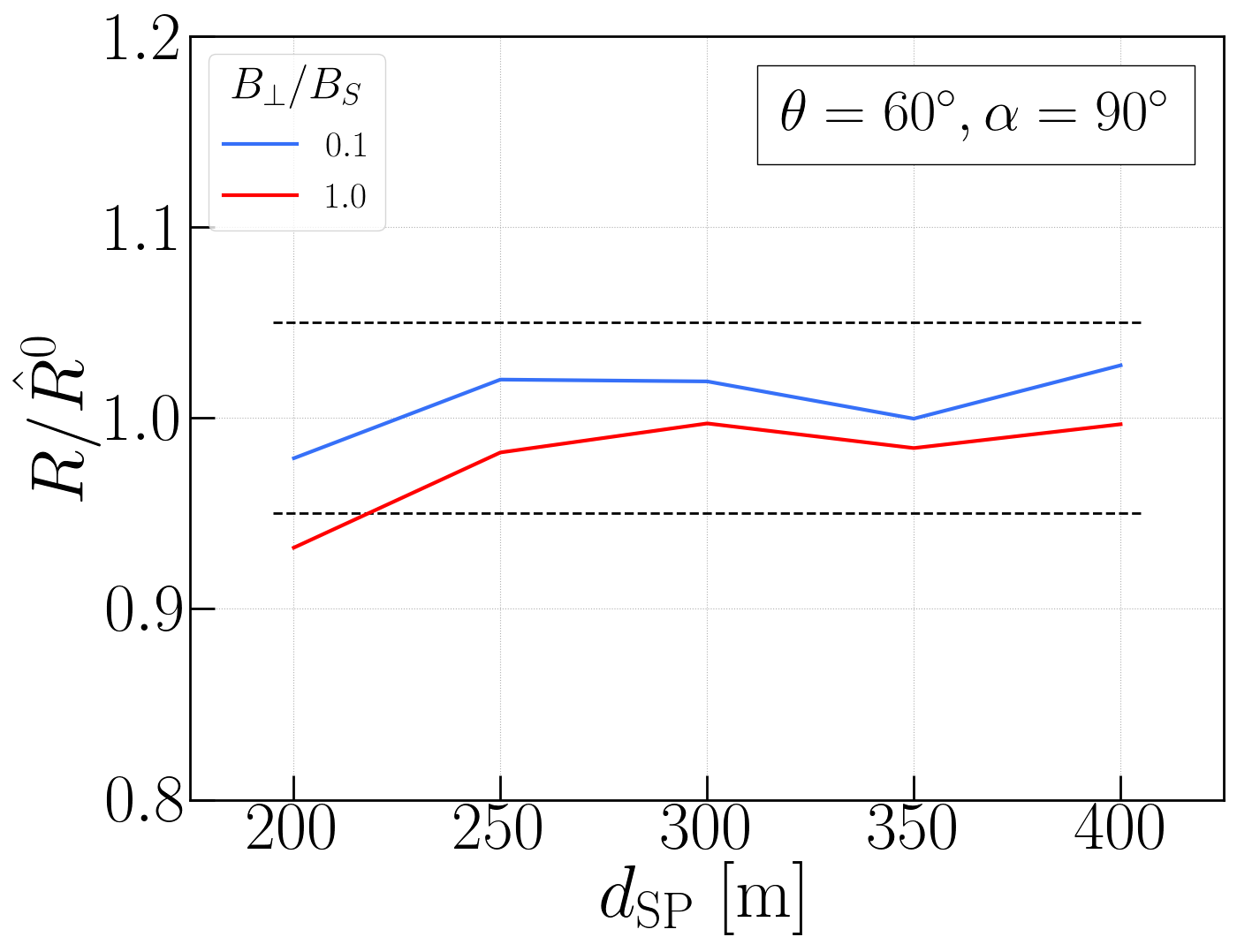}
		\caption{
			Top: Shower plane (SP) orthogonal to the air shower direction. 
			T1$_{\mathrm{SP}}$ and T2$_{\mathrm{SP}}$ are the positions of a pair of telescopes projected from the horizontal to the shower plane. 
			The distance between both telescopes is~$d_{\mathrm{SP}}$. 
			$\vec{B}_{\perp}$ is the component of the Earth's magnetic field vector perpendicular to the air shower direction. 
			The angle between $\vec{B}_{\perp}$ and the line joining two telescopes in the shower plane is denoted $\alpha$. 
			Bottom: The ratio $R / \Fnom{}$ for the simulated SST pairwise trigger rates ($R$) assuming $B = B_{\mathrm{S}}$, where $B_{\mathrm{S}}$ is the modulus of the geomagnetic field vector at the CTA South site. 
			The rate estimates ($\Fnom{}(\theta, d_{\mathrm{SP}})$) were obtained from the fit of geometrical dependencies for $B \approx 0$ (see also Fig.~\protect\ref{Fig:Geometry}). 
			Shown are different scenarios for $B_{\perp} / B_{\mathrm{S}}$ (colors) and the angle $\alpha = 0^{\circ}$ (left) and $90^{\circ}$ (right). 
			Dashed lines denote a $5\%$ difference between  $R$ and $\Fnom{}$.
		}
		\label{Fig:Bfield}
	\end{center}
	\vspace{-0.7cm}
\end{figure}

The term $\Fnom{ij}$ assures that the CTC is independent of the observational conditions, represented by the layouts of telescope pairs and the zenith and azimuth angles of incident air showers. 
Fig.~\ref{Fig:Geometry} illustrates the relationship between the pairwise trigger rate and the distance $d_{\mathrm{SP}}$ between the telescope positions projected on the plane orthogonal to the air shower direction (shower plane, see the top panel in Fig.~\ref{Fig:Bfield}). 
Stereo trigger rates were obtained from MC simulations for pairs constituting telescopes of the same type.  Negligible strength of the magnetic field was assumed ($B \approx 0.001 \mu$T) to examine only the geometrical effects. 
Three designs of different-sized telescopes foreseen at the CTA South were assumed.  
The impact of the zenith angle ($\theta$) of observations is depicted for $\theta = 0^{\circ}$ (left panel) and $60^{\circ}$ (right). 
Dashed lines denote the approximation of the trigger rate by a continuous function $\Fnom{ij} (\theta, d_{\mathrm{SP}})$ recovered from the fit of the MC data separately for each telescope class and for vanishing magnetic field, see e.g.~\mbox{\cite{Stefanik:2019a}}.

In addition to the hardware and observation configurations, the local geomagnetic field ($\vec{B}$) also affects telescope trigger rates, see the bottom panels in Fig.~\ref{Fig:Bfield}. 
As an example, we depict pairwise trigger rates ($R$) of the SSTs which were obtained from simulations with $B_{\mathrm{S}} = |\vec{B}_{\mathrm{S}}|$ corresponding to the CTA South site. 
The evolution of the ratio $R / \Fnom{} (\theta, d_{\mathrm{SP}})$ with $d_{\mathrm{SP}}$ is shown for different relative sizes of the component of the magnetic field vector perpendicular to the air shower direction ($B_{\perp} / B_{\mathrm{S}}$, see the top panel in Fig.~\ref{Fig:Bfield}). 
We explore also the impact of the angle $\alpha$ between $\vec{B}_{\perp}$ and the inter-telescope distance projected into the shower plane. 
For $\alpha = 0^{\circ}$ (left) and $90^{\circ}$ (right), charged particles in an air shower are deflected predominantly in the direction orthogonal and parallel to the line joining the telescopes, respectively. 
Due to the computational restrictions, the statistical uncertainties of the shown trigger rates reach up to $20\%$, depending on the inter-telescope distance. 
The results of Fig.~\ref{Fig:Bfield} are regarded as estimates on the systematic uncertainties of the CTC. 

If uncorrected, the mis-reconstruction of the pairwise trigger rates at the CTA South  can be as much as $10\%$ (red lines) while the attainable accuracy is within $5\%$ for pointing directions roughly parallel with $\vec{B}_{\mathrm{S}}$ (blue lines). 
It is worth noting that the modulus of the magnetic field vector at the CTA North is $B_{\mathrm{N}} \approx 1.7 B_{\mathrm{S}}$. 
While all pointing directions at the CTA South with $\theta \leq 60^{\circ}$ are equivalent to $B_{\perp} / B_{\mathrm{N}} < 0.6$, only $\sim 17\%$ of configurations at the CTA North satisfy this inequality. 
The accuracy of the CTC for uncorrected effects of the magnetic field is expected to be better at the southern observatory as larger values of $B_{\perp}$ are assumed to result in larger bias on the trigger rate.

% ******************************************************************************************
% APPLICATION OF THE CTC AT THE CTA SOUTH
% ******************************************************************************************
\vspace{-0.1cm}
\section{Application of the CTC at the CTA South}
\label{Sec:Application}
\vspace{-0.1cm}

The feasibility of the CTC was tested using simulations of proton-induced air showers observed by the full CTA South array consisting of 4 LSTs, 25 MSTs and 70 SSTs~\cite{Maier:2019}. 
The observation altitude and magnetic field were set correspondingly to the CTA South site. 
Zenith and azimuth angles were chosen as $60^{\circ}$ and $180^{\circ}$, respectively, in order to explore the performance of the method for a small influence of the geomagnetic field ($B_{\perp} / B_{\mathrm{S}} = 0.1$). 
Telescopes were assigned randomly degraded optical throughput efficiencies from the normal distribution $\mathcal{N} (0.7, 0.1)$. 

% ******************************************************************************************
% Atmospheric monitoring
% ******************************************************************************************
\vspace{-0.1cm}
\subsection{Atmospheric monitoring}
\label{Sec:Monitoring}

We test the performance of the CTC for the monitoring of changes in the atmospheric transmission due to different concentrations of aerosols. 
We assume four different models of aerosol optical depth ($\tau$) as provided by the MODTRAN software~\cite{Berk:2014}. 
The used atmospheric profiles simulate no aerosol extinction, extinction due to aerosols in the boundary layer close to the ground or extinction higher in the troposphere but below the production height of VHE air showers. 
We~calculate the atmospheric transparency $T$ corresponding to these profiles as the median value of $\exp(-\tau (\lambda))$ weighted by the mirror reflectivity and the quantum efficiency of the camera photo-sensors which are both functions of the wavelength of incident Cherenkov light ($\lambda$). 
Here, $\tau$ is the aerosol optical depth in a vertical air column\footnote{The zenith angle dependence of the optical depth is accounted for by the normalization $\Fnom{}$ in Eq.\eqref{Eq:CTC}.} from the observation level up to the height of 18~km above sea level, i.e. above the emission altitudes of the majority of Cherenkov photons from air showers. 
The transparency is normalized so that $T = 1$ corresponds to no aerosol attenuation.
A single model of the molecular density of the atmosphere is assumed as it has been shown that the CTC is not sensitive to the seasonal variations of the local density profile~\cite{Stefanik:2019b}.

Out of the whole array we choose 6 LST, 72 MST and 100 SST pairs of telescopes of the same type with inter-telescope distances $<300$~m. 
For each of the telescope pairs, an estimate of the pairwise trigger rate ($\Fnom{ij}$) is obtained from fits of MC trigger rates in a variety of geometrical configurations. 
For details of the used fit functions see~\cite{Stefanik:2019a}. 

\begin{figure}[!t]
	\begin{center}
		\includegraphics[width=0.42\columnwidth]{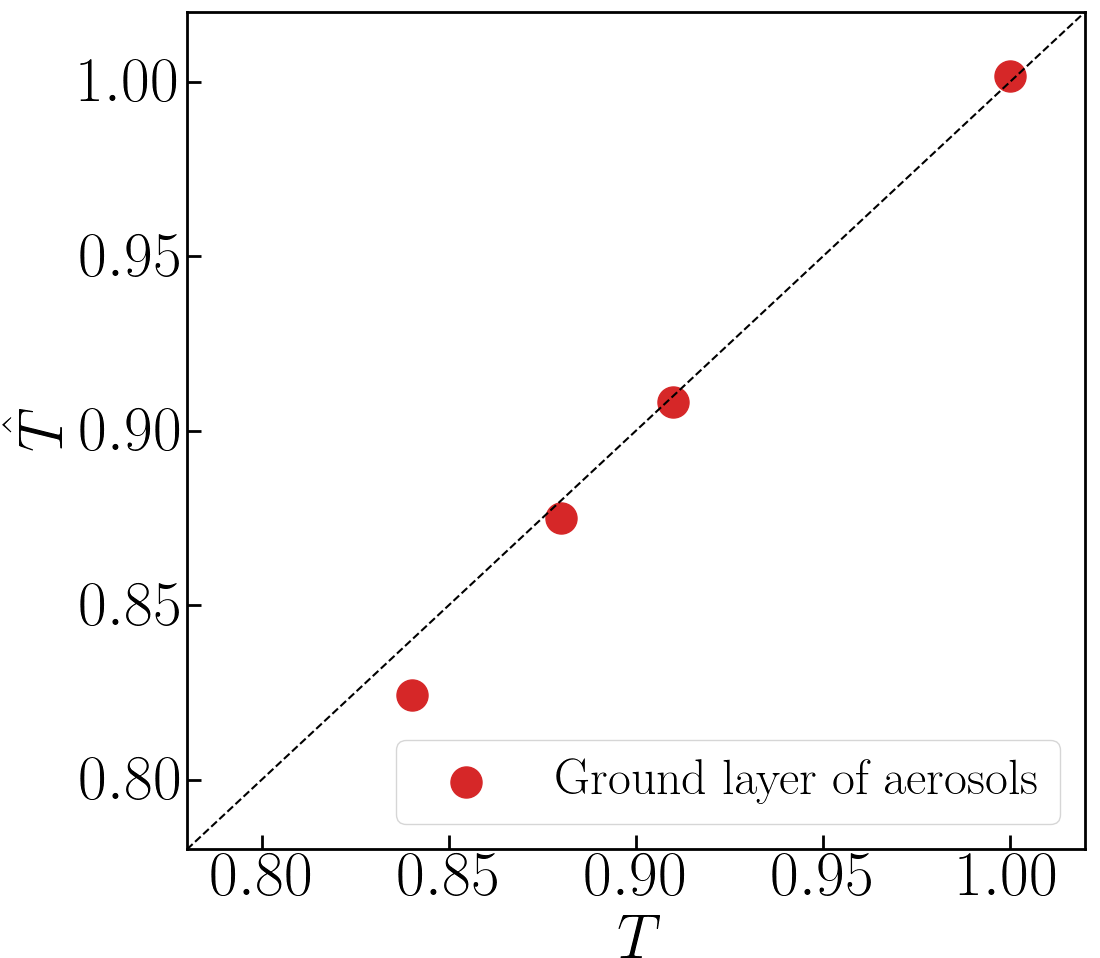}
		\hspace{0.05\columnwidth}
		\includegraphics[width=0.42\columnwidth]{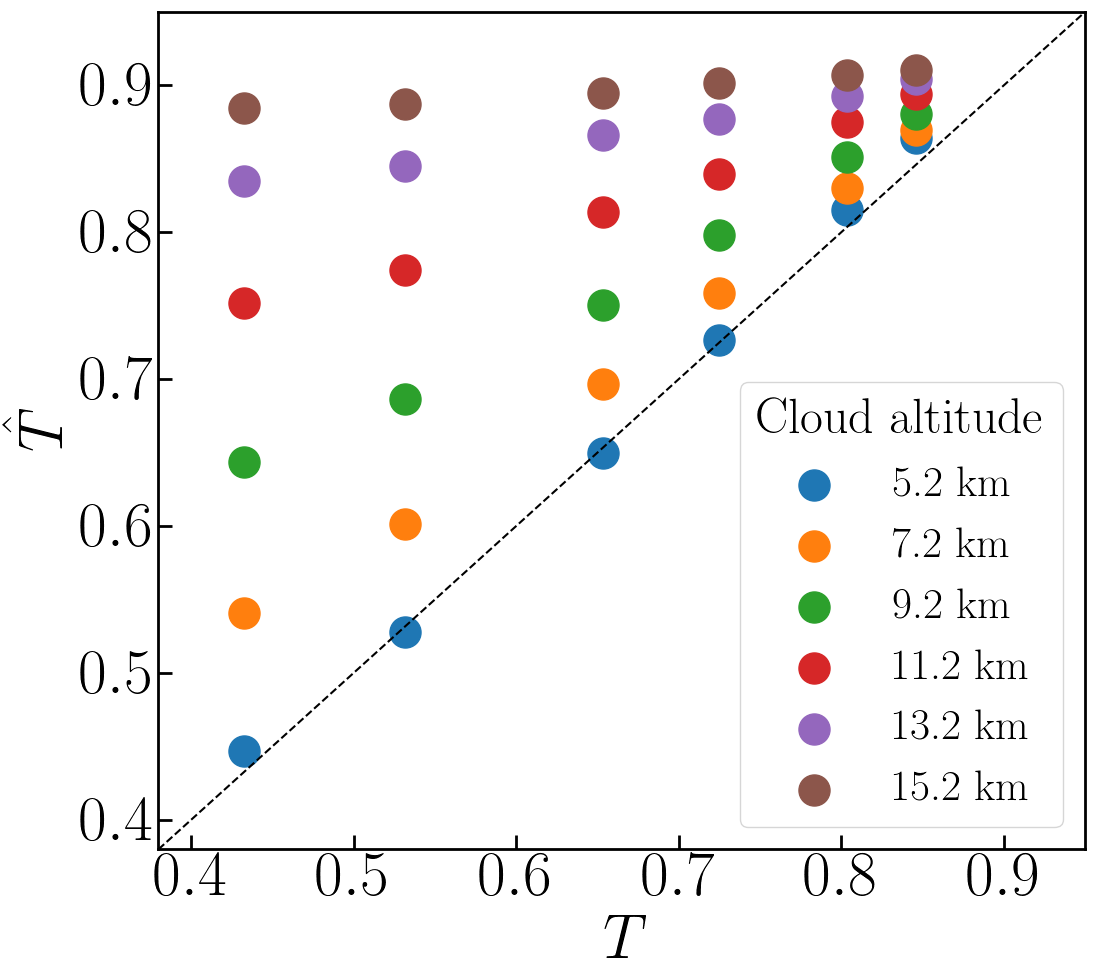}
	
		\caption{
			The Cherenkov transparency coefficient ($\tij{}$) retrieved from the MC trigger rates of the full CTA South array. 
			Left: the CTC for the simulations assuming four different concentrations of aerosols in the ground layer below the altitudes of air showers. 
			Right: the CTC for clouds of various aerosol optical depths simulated at different altitudes (colors). 
			The reference MC atmospheric transparency ($T$) is calculated from the ground level (2.2~km) up to a height of 18~km a.s.l. 
			Dashed lines represent the equality $\tij{} = T$. 
		}
		\label{Fig:Monitoring}
	\end{center}
	\vspace{-0.8cm}
\end{figure}

The CTCs are calculated according to Eq.\eqref{Eq:CTC} as averages of pairwise transparency estimates over all selected 178 telescope pairs. 
In the left panel in Fig.~\ref{Fig:Monitoring}, the CTCs ($\tij{}$) are compared with the MC atmospheric transparency ($T$) for the models of the ground layer aerosols. 
The CTC is sensitive to the increased atmospheric attenuation and agrees with the input transparency within~$1\%$. 
However, it has to be emphasized that the optical throughput ($\varepsilon$) used for the calculation is assumed to be perfectly known and is substituted by the throughput set in MC simulations. 
In~reality, the optical throughput efficiencies will be provided by other calibration methods~\cite{Maccarone:2017}. 
It~is expected that the resolution of the transparency estimates will worsen according to the precision of the reconstructed optical throughput, e.g. $\sim 4-5\%$ when using muon rings~\cite{Gaug:2019}. 
A further contribution to the systematic uncertainty of the CTC stems from the influence of the geomagnetic field which is more significant in observations conducted under pointing directions nearly perpendicular to the field vector (Fig.~\ref{Fig:Bfield}). 
Note that this uncertainty may be reduced through detailed MC simulations of the pairwise trigger rates in different bins of $B_{\perp} / B_{\mathrm{S}}$. 

In addition to the ground layer aerosols, we inspect the CTC performance under the presence of clouds. 
The MODTRAN input for our simulations assumes clouds with a uniform structure and thickness of 1~km. 
Altitudes of cloud bases have been chosen to $5.2-15.2$~km~a.s.l. 
Aerosol optical depths of clouds were chosen in the range $0.05-0.7$.  
Besides the cloud specifications, all used models assume the same atmospheric profile with desert-like extinction due to aerosols. 
The atmospheric transparency is again calculated from the total aerosol optical depth below 18~km. 

The CTCs corresponding to each cloud configuration are shown in the right panel in Fig.~\ref{Fig:Monitoring}. 
Atmospheric transparency estimates agree with the simulated values for low lying clouds which are completely below the maxima of Cherenkov radiation from VHE air showers ($\sim5$~km a.s.l.). 
With increasing altitude, clouds obstruct less Cherenkov light and the atmospheric transparency is overestimated. 
At 15~km, the CTC is almost insensitive to changes of the cloud optical properties. 

For aerosol attenuation close to the ground, the CTC is a robust measure of the atmospheric influence on the energy reconstruction within the IACT technique. 
However, the bias in the reconstructed energy of primary $\gamma$--rays is height-dependent for absorbers at higher altitudes. 
While the CTC is sensitive to changes in the cloud altitude, neither the altitude nor the optical depth of the cloud can be ascertained. 
For higher lying clouds, the CTC can be only used as a complementary method to approaches capable of assessing the altitude of an absorber, e.g.~the Raman~LIDAR~\cite{Vasileiadis:2019}.

% ******************************************************************************************
% Cross-calibration of telescope responses
% ******************************************************************************************
\vspace{-0.1cm}
\subsection{Cross-calibration of telescope responses}
\label{Sec:Cross_calibration}

\begin{figure}[!t]
	\begin{center}
		\includegraphics[width=0.45\columnwidth]{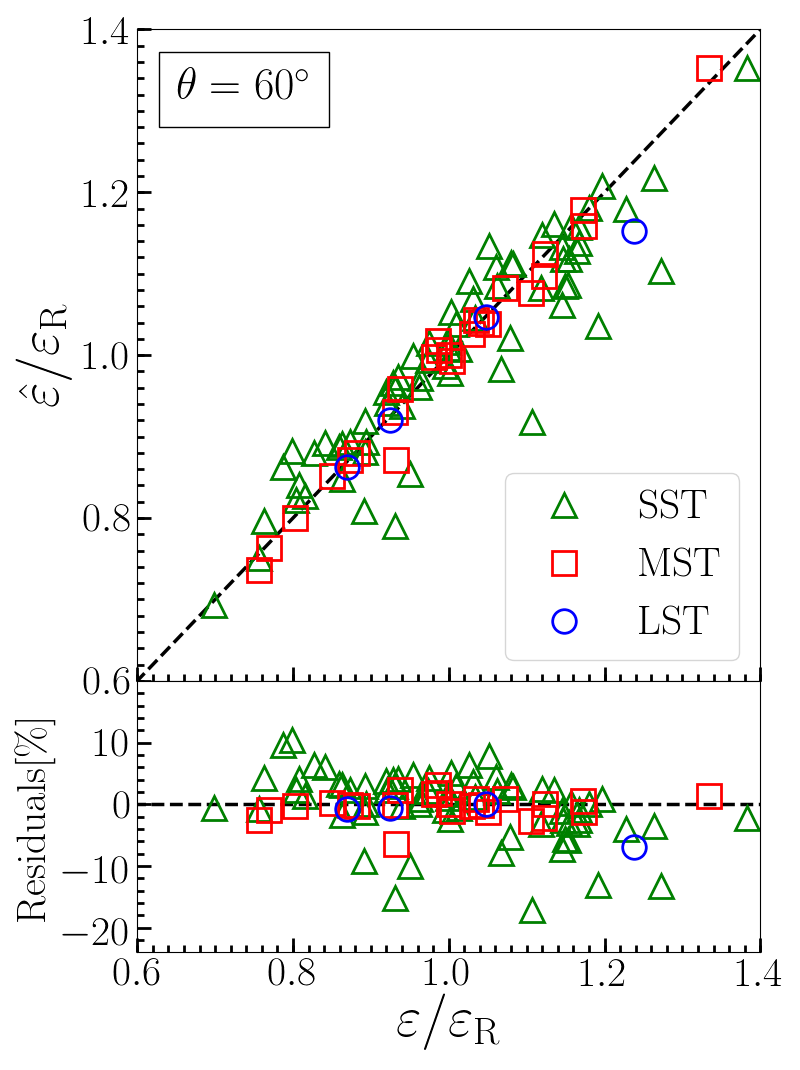}
		\caption{
			Relative estimates of the optical throughput ($\hat{\varepsilon}_{i} / \varepsilon_{\mathrm{R}}$) from cross-calibration of telescopes in the full CTA South array are compared to the efficiencies assumed in MC simulations ($\varepsilon_{i} / \varepsilon_{\mathrm{R}}$).	
			The results are shown relative to the optical throughput of a reference telescope which was not varied during the optimization outlined in Section~\protect\ref{Sec:CTC}.
			Percentage residuals are shown below. 
			Colors denote different telescope classes.
		}
		\label{Fig:Cross_calibration}
	\end{center}
	\vspace{-0.7cm}
\end{figure}

In another application of our method, we assume that the optical throughput of telescopes is not known and needs to be estimated. 
To this end, we calculate pairwise transparency estimates ($\tnom{ij}$, see Section~\ref{Sec:CTC}) for the case of nominal optical throughput $\left( \varepsilon = 1 \right)$. 
The true atmospheric transparency is not known but presumed to be uniform across the whole telescope array.\footnote{Given the extent of the array, this requirement might not always be met. In such case the inter-calibration may be performed within smaller groups of telescopes for which variations of atmospheric conditions are assumed to be small.}  
In~each sub-system of telescopes of class $C$, where $C \in \left\lbrace \text{LST, MST, SST} \right\rbrace$, we select arbitrarily one reference telescope to fix the scale of the optical throughput ($\varepsilon_{\mathrm{R}}^{C}$) and optimize the efficiencies of the remaining telescopes and the atmospheric transparency according to Eq.\eqref{Eq:chi2}. 
The output of the minimization are three sets of inter-calibrated optical throughput efficiencies $\hat{\varepsilon}_{i}^{C} / \varepsilon_{\mathrm{R}}^{C}$. 
The ambiguity between the reference efficiencies of individual classes is then removed by the cross-calibration outlined in Section~\ref{Sec:CTC}. 

The reconstructed optical throughput efficiencies are drawn against those assumed in MC simulations in Fig.~\ref{Fig:Cross_calibration}. 
In the shown example, the resolution of the cross-calibration is better than $4\%$ for the LSTs and MSTs and $7\%$ for the SSTs. 
Outliers in the set of reconstructed SST efficiencies correspond to the telescopes at the edge of the array which have fewer other telescopes in their vicinity. 
The inter-telescope distances for pairs constituting edge telescopes are larger than for instruments near the core of the array, leading to worse statistics for pairwise trigger. 

We note that the method may perform differently under adverse observation conditions, e.g. elevated levels of the night-sky background or $B_{\perp} / B_{\mathrm{S}} \approx 1$ (unless accounted for). 
Carried out in a relative way, the cross-calibration using the CTC is then complementary to the absolute calibration using muon ring images with a precision of $4\%$~\cite{Gaug:2019}.

% ******************************************************************************************
% CONCLUSIONS
% ******************************************************************************************
\vspace{-0.2cm}
\section{Conclusions}
\label{Sec:Conclusions}

We have demonstrated the feasibility of the amended Cherenkov transparency coefficient~\cite{Stefanik:2019a} for the atmospheric and array calibration at the southern array of the CTA observatory. 
Utilizing the output from normal data taking, the CTC reproduces correctly the evolution of the atmospheric transmission to Cherenkov photons for the aerosol layers below the production height of VHE air showers. 
While being in good agreement with the input transparency within the Monte Carlo study, systematic uncertainties of $5-8\%$ can be expected, dominated by the impact of the geomagnetic field and the precision of the optical throughput estimates. 
As to the monitoring of the optical throughput using the CTC concept, a relative calibration of the telescope array has been achieved with a resolution of $4-7\%$, depending on the telescope class.
The influence of the geomagnetic field has been shown to be a less limiting factor compared to the usage at the CTA North. 

For the first time, we investigated the applicability of the CTC for the CTA sub-array of small-size telescopes, the SST-1M design in particular chosen from three SST proposals. 
Our simulation results are not expected to strongly depend on this choice because the CTC viability for the SSTs depends mainly on the size of the primary mirror dish and the telescope array layout. 
The CTC concept has been shown to be a feasible means of the inter-calibration for the SSTs, albeit with worse precision than for the larger telescope classes (Fig.~\ref{Fig:Cross_calibration}) due to the sparser spacing of the~SSTs.

The focus of the future study will be on the stability of the CTC under different intensities of the night-sky background and variable trigger thresholds of the LSTs. 
The verification of the CTC against other calibration approaches is foreseen with the installation of the first CTA telescopes.

%***********************************************************************************
% ACKNOWLEDGEMENTS
%***********************************************************************************

\vspace{-0.4cm}
\acknowledgments
\vspace{-0.2cm}
This work was conducted in the context of the CTA Central Calibration Facilities. 
We gratefully acknowledge financial support from the agencies and organizations listed here: http://www.cta-observatory.org/consortium\_acknowledgments. 
We are also grateful for the support by the grants LTT17006 and LM2015046 of the Ministry of Education, Youth and Sports of the Czech Republic.

% ******************************************************************************************
% BIBLIOGRAPHY
% ******************************************************************************************

% ******************************************************************************************
% ******************************************************************************************
% ******************************************************************************************
% END OF DOCUMENT
% ******************************************************************************************
% ******************************************************************************************
% ******************************************************************************************

\end{document}